\documentclass[aip,reprint,nofootinbib]{revtex4-2}
\usepackage[utf8]{inputenc}
\usepackage{color}
\usepackage{amssymb,amsmath,bm,mathtools,relsize}
\usepackage{mathrsfs}
\usepackage{url}
\usepackage{graphicx}
\usepackage{tikz}

\newcommand{\uu}{\mathbf{u}}
\newcommand{\vv}{\mathbf{v}}

\begin{document}

\title{Koopman analysis of the periodic Korteweg-de Vries equation}
\author{Jeremy P Parker}
\email[Author to whom correspondence should be addressed: ]{jeremy.parker@epfl.ch}
\affiliation{Emergent Complexity in Physical Systems Laboratory (ECPS), \'Ecole Polytechnique F\'ed\'erale de Lausanne, Lausanne, Switzerland}
\author{Claire Valva}
\affiliation{Courant Institute of Mathematical Sciences, New York University, New York, New York, USA}

\begin{abstract}
    The eigenspectrum of the Koopman operator enables the decomposition of nonlinear dynamics into a sum of nonlinear functions of the state space with purely exponential and sinusoidal time dependence.
    For a limited number of dynamical systems, it is possible to find these Koopman eigenfunctions exactly and analytically.
    Here, this is done for the Korteweg-de Vries equation on a periodic interval, using the periodic inverse scattering transform and some concepts of algebraic geometry.
    To the authors' knowledge, this is the first complete Koopman analysis of a partial differential equation which does not have a trivial global attractor.
    The results are shown to match the frequencies computed by the data-driven method of dynamic mode decomposition (DMD). We demonstrate that in general DMD gives a large number of eigenvalues near the imaginary axis, and show how these should be interpretted in this setting.
\end{abstract}

\maketitle

\begin{quotation}
Dynamic mode decomposition (DMD) is a widely used computational method for analysing spatiotemporal data from experiments, observations and numerics. The connection to the mathematical Koopman operator means that we can understand the behaviour of DMD by analytically applying the Koopman operator to integrable partial differential equations. One non-trivial example is the Korteweg-de Vries equation on a periodic domain, which admits both wavelike and soliton solutions, and can be solved analytically via the inverse scattering method.
\end{quotation}

\section{Introduction}
The Koopman operator was introduced by \citet{koopman1931hamiltonian} to describe the nonlinear behaviour of a dynamical system as the linear evolution of nonlinear observables of that system. It is well known that a finite dimensional nonlinear system can be converted into an infinite dimensional linear system; Koopman analysis extends this to infinite dimensional nonlinear systems. Recent interest in the Koopman operator was initiated by Igor Mezi\'c \citep{Mezic2005,Mezic2013}, and it has become popular for its close connection to the computational method of dynamic mode decomposition.

DMD was originally invented by \citet{Schmid2010} as a method for simultaneously extracting the important spatial and temporal features of a timeseries. It has been successfully applied to a wide array of numerical, observational and experimental data, most notably in fluid dynamics \citep{schmid2011applications,kutz2016dynamic,schmid_dynamic_2022}. 
Under certain conditions \citep{Rowley2009}, the results of DMD can be seen as a numerical approximation to the Koopman modes and eigenvalues of the underlying dynamical system, and so understanding the Koopman operator aids interpretation of the results of DMD.
In particular, to understand the spatial patterns called DMD modes, it would be helpful to have analytic results on nonlinear partial differential equations (PDEs).

Several authors \citep{kutz2018applied,page2018koopman,balabane2021koopman} have successfully performed Koopman analysis on the Burgers equation, a dissipative nonlinear PDE which can be tranformed into the linear heat equation with a suitable change of variables.
\citet{nakao2020spectral} considered the Burgers equation as well as a non-trivial transformation of it to the phase-diffusion equation.
Though insightful, these PDEs admit only a single steady global attractor, and are strongly dissipative, which precludes a lot of interesting nonlinear behaviour relevant to applying DMD to situations involving sustained waves, a common use-case.

\citet{parker2020koopman} considered the Korteweg-de Vries (KdV) equation, an integrable partial differential equation of one variable. This Hamiltonian dynamical system behaves very differently from the dissipative systems mentioned above. In that work, some Koopman eigenfunctions were found for soliton solutions of the KdV equation on the real line which excluded a large class of solutions; in particular, this excluded the spatially periodic solutions which are the natural choice for computer simulations of solitons. It was argued that the simplest periodic solutions, cnoidal waves, give purely imaginary Koopman eigenvalues, in sharp contrast to the isolated solitons, which have purely real Koopman eigenvalues, despite being a naturally limiting case of the former.

A periodic domain is the natural setting for numerical solutions of 1-dimensional PDEs like the KdV equation.
Indeed, early pioneering numerical simulations of the KdV equation on a periodic domain \citep{zabusky1965interaction} were used to shed light on the famous Fermi-Pasta-Ulam-Tsingou problem\citep{fermi1955studies}, in which the initial state of a system recurs to arbitrary precision after complex nonlinear dynamics. It was later proven\citep{lax1976almost} that this is because the solutions lie on quasi-periodic invariant tori. We will show that it is possible to use DMD to determine the underlying frequencies. In fact, when the dynamics are confined to an invariant torus, DMD is equivalent to a Fourier transform in time, though it generalises when other modes grow or decay.

In the present paper, we study a particular but very general class of solutions of the KdV equation on a periodic interval, for which we are able to define Koopman eigenfunctions. These eigenfunctions require the evaluation of contour integrals on Riemann surfaces which must be performed numerically in all but the simplest of cases. This method is not, therefore, recommended as a general approach for nonlinear PDEs, where DMD could easily be applied. However, the semi-formal mathematical treatment presented here gives instructive results: we will see that in this case, Koopman eigenvalues necessary for the decomposition of the state of the system densely fill the imaginary axis, and so the results of DMD are subtle to interpret.
The paper proceeds as follows: in section \ref{sec:kdv} the KdV equation is introduced and its history and significance briefly described; in section \ref{sec:koopman} the Koopman operator and its spectrum are defined; in section \ref{sec:eigenfunctions} we then derive Koopman eigenfunctions for the KdV equation; section \ref{sec:results} presents the results of applying these to an example solution, which is compared to the numerical results of DMD in section \ref{sec:dmd}. Concluding remarks are given in section \ref{sec:discussion}.

\section{The KdV equation}
\label{sec:kdv}

The KdV equation\footnote{Many different conventions are employed in the literature, and notably solitons can be either positive or negative depending on this choice. Here we follow \citet{belokolos1994algebro}.}
\begin{align}
    \begin{split}
    \label{eq:kdv}
            &4\partial_t u(x,t) = 6u(x,t)\,\partial_x u(x,t) + \partial_x^3 u(x,t), \\ &u(x,0)=u_0(x),
    \end{split}
\end{align}
was derived by \citet{korteweg1895xli} to describe the weakly nonlinear evolution of shallow water waves propagating in one direction. The pioneering computational results of \citet{zabusky1965interaction} demonstrated the existence of soliton solutions when this equation is solved on a finite periodic domain. For an infinite domain, the celebrated inverse scattering method of \citet{gardner1967method} gives a straightforward procedure to solve the equation, and provides intuitive interpretations for the existence of solitons as conserved quantities. Analytical results on a periodic interval have proven much more complicated, despite the early computation successes and the well known cnoidal wave solution.

In addition to shallow water waves, the KdV equation naturally arises from weakly nonlinear theory in many physically relevant flows \citep{benney1966,peregrine1966calculations,karpman1975non}. Our interest however derives from the fact that the equation admits complex, nonlinear but non-chaotic solutions amenable to analytic treatment. Unlike other PDEs for which Koopman spectra have been derived, it is a Hamiltonian system with an infinite number of conserved quantities, rather than having one unique attractor.

The solution to (\ref{eq:kdv}) is  well-posed on  $\mathbb{T}=\mathbb{R}/2\pi\mathbb{Z}$, the periodic domain of length $2\pi$, for initial conditions in Sobolev spaces $H^s(\mathbb{T},\mathbb{R})$ with $s\geq -1$ \citep{Kappeler2006}, with a well-defined evolution operator $\mathscr{S}_t:H^s(\mathbb{T},\mathbb{R})\to H^s(\mathbb{T},\mathbb{R})$ for each time $t$.
A single periodic wave of sufficient amplitude breaks down into a spatially periodic solution, with quasi-periodic behaviour in time. That is to say, solutions lie on an invariant torus, as is usual for integrable Hamiltonian systems, and these invariant tori foliate phase space. In fact, any solution of (\ref{eq:kdv}) can be approximated to arbitrary precision as an invariant $M$-torus \citep{lax1976almost}, $M\in\mathbb{N}$. These are the so-called `finite gap' solutions, which will be the focus of the present study. Practically, the evolution of an arbitrary initial condition can be approximated by truncating the scattering data at a judiciously chosen $M$ \citep{christov2012hidden}.

Let us therefore define our solution space $\Omega_M$ to be the subset of $L^2(\mathbb{T},\mathbb{R})$ for which there are $M$-gap solutions (this terminology should become clearer in section \ref{sec:eigenfunctions}). This is a well-posed invariant subspace. For convergent Koopman decompositions it will be necessary to further restrict this space, in a manner analogous to that of \citet{balabane2021koopman}, in section \ref{sec:results}. 

It will be useful later to make the so-called Hirota transformation $u\mapsto\vartheta$ defined by \citep{hirota2004direct}
\begin{equation}
\label{eq:hirota}
    u(x,t) = 2\partial_x^2 \log{\vartheta(x,t)},
\end{equation}
so that (\ref{eq:kdv}) becomes
\begin{equation}
    4\vartheta\partial_x\partial_t\vartheta-4\partial_x\vartheta\partial_t\vartheta-3(\partial_x^2\vartheta)^2+4\partial_x\vartheta\partial_x^3\vartheta-\vartheta\partial_x^4\vartheta=0.
\end{equation}
This is analogous to the Cole-Hopf transformation exploited by previous authors for the Burgers equation \citep{kutz2018applied,page2018koopman,balabane2021koopman}, though in this case the transformation seems at first glance to have made the equation more complicated. The utility comes from the fact that $\vartheta$ can be expressed as a Riemann theta function, as explained in section \ref{sec:eigenfunctions}.

\section{The Koopman operator}
\label{sec:koopman}

Let $\mathcal{O}_M$ be the space of continuous maps $\Omega_M\to\mathbb{C}$, which are called observables of the system.
The Koopman operator, a composition operator for dynamical systems, is defined for each $t\geq0$ by
\begin{align}
\begin{split}
\mathcal{K}^t_M:\mathcal{O}_M&\to\mathcal{O}_M,\\
    [\mathcal{K}^t_M \phi]\left(u_0\right)&=\phi\left(u(\cdot,t)\right),
\end{split}
\end{align}
for any observable $\phi\in\mathcal{O}_M$, where the evolution of $u(x,t)$ is governed by (\ref{eq:kdv}). The Koopman operator is a linear operator, amenable to spectral theory.
The \emph{Koopman eigenvalues} $\lambda\in\mathbb{C}$ and \emph{Koopman eigenfunctions} $\varphi\in\mathcal{O}_M$ satisfy
\begin{equation}
    [\mathcal{K}^t_M \varphi]\left(u_0\right)=e^{\lambda t}\varphi\left(u_0\right)
\end{equation}
or equivalently
\begin{equation}
    \varphi\left(u(\cdot,t)\right)=e^{\lambda t}\varphi\left(u_0\right).
\end{equation}
A simple example of a Koopman eigenfunction would be any conserved quantity of the dynamics, with eigenvalue $\lambda=0$. More generally, they are any observable for which the temporal behaviour is purely (complex) exponential as the state evolves. Since the system we study is Hamiltonian, we expect only purely imaginary eigenvalues $\lambda=i\omega$, giving purely oscillatory behaviour. In certain circumstances, it may be possible that the Koopman eigenfunctions form a basis for $\mathcal{O}_M$, in which case we can decompose all other observables as a sum over Koopman eigenfunctions. In particular, we are interested in whether it is possible to write the state of the system $u$ as a convergent sum
\begin{equation}
    u(x,t) = \sum_{\nu} c_\nu(x)\,\varphi_\nu(u_0)\,e^{i\omega_\nu t}.
\end{equation}
Here the $c_\nu:\mathbb{T}\to\mathbb{C}$ are called \emph{Koopman modes}, which encode spatial information for each eigenvalue and are independent of the particular choice of initial condition $u_0$, whose contribution is included in the value of $\varphi_\nu(u_0)$. If this is possible, it means that the dynamics of (\ref{eq:kdv}) can be decomposed as a sum over nonlinear functions whose temporal behaviour is purely oscillatory in time.

\section{Koopman eigenfunctions of the KdV equation}
\label{sec:eigenfunctions}

It would take a whole textbook to fully explain the periodic inverse scattering transform. We refer readers to the textbooks by \citet{novikov1984theory}, \citet{belokolos1994algebro} and \citet{osborne2010nonlinear} for accessible introductions, including the necessary background of Riemann surfaces and theta functions, though note the differing notations and conventions between these (we follow the notation of \citet{belokolos1994algebro}). Here we give a summary of the relevant results for the KdV equation which are implemented in the Mathematica notebook given in the supplementary materials.

\begin{figure}
    \centering
    \includegraphics[width=\columnwidth]{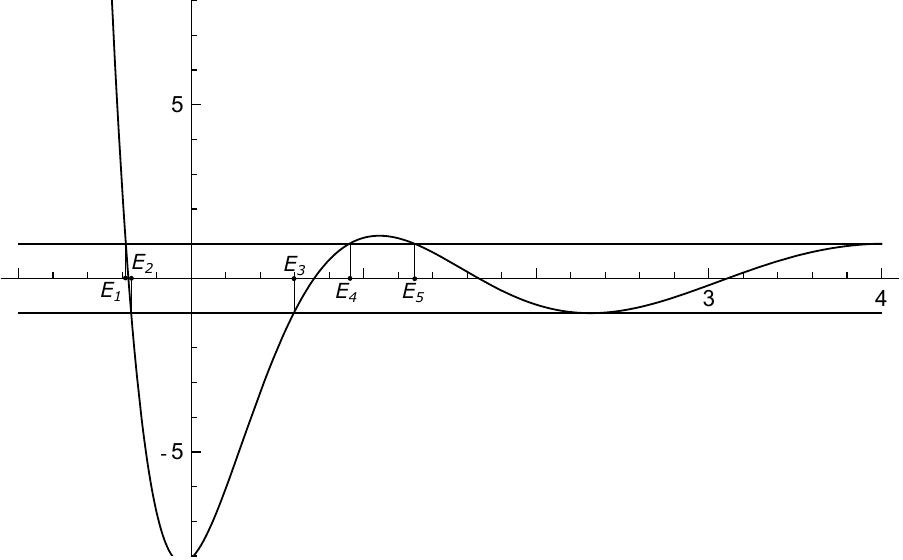}
    \caption{$F(\lambda)$, half the trace of the monodromy matrix, against real $\lambda$, for $u_0(x)=\sin{x}$. Allowable values of $\lambda$ for a bounded eigenfunction are when $|F(\lambda)|\leq 1$. The forbidden regions $(-\infty,E_1)$, $(E_2,E_3)$, \dots are called gaps.}
    \label{fig:monodromy}
\end{figure}

One of the key results in the solution of the KdV equation was the discovery of a Lax pair \citep{lax1968integrals}: a pair of linear operators $L(u),A(u):L^2(\mathbb{R})\to L^2(\mathbb{R})$ such that
\begin{equation}
    \frac{dL}{dt} = L(t)\circ A(t) - A(t) \circ L(t),
    \label{eq:laxpair}
\end{equation}
where $L(t):=L(u(\cdot,t))$ etc.
In the case of the KdV equation (\ref{eq:kdv}),
\begin{align}
    L(u) &= -\frac{d^2}{dx^2} - u,\label{eq:L}\\
    A(u) &= \frac{d^3}{dx^3}+\frac{3}{4}\left(u\frac{d}{dx}+\frac{d}{dx}u\right).\label{eq:A}
\end{align}
The operator (\ref{eq:A}) is skew-adjoint.
The operator (\ref{eq:L}) is the well known self-adjoint Schr\"odinger operator, with potential $u$.
From (\ref{eq:laxpair}) it can be shown that the spectrum of $L$ is independent of $t$, when $u(x,t)$ satisfies (\ref{eq:kdv}). Finding eigenvalues $\lambda\in\mathbb{R}$ and eigenfunctions $\psi\in C^{\infty}$ reduces to the Sturm-Liouville problem
\begin{equation}
\label{eq:eigenvalueproblem}
    \psi''+u\psi+\lambda \psi = 0.    
\end{equation}
In the case of a periodic potential $u(x+2\pi)=u(x)$, equation (\ref{eq:L}) is known as Hill's operator and has been widely studied\citep{magnus2013hill}. The admissible eigenvalues $\lambda$ for a bounded eigenfunction $\psi$ reside in intervals $[E_1,E_2]$, $[E_3,E_4]$, \dots, where $-\infty<E_1<E_2\leq E_3<E_4\leq E_5<\dots$. Outside these regions, only unbounded solutions are possible, and these are termed forbidden gaps. The $E_k$ are the values of $\lambda$ for which $F(\lambda)$, defined as half the trace of the monodromy matrix of (\ref{eq:eigenvalueproblem}), is $\pm 1$ \citep{magnus2013hill} (see figure \ref{fig:monodromy}).
Though the monodromy matrix is not invariant under the dynamics (\ref{eq:kdv}), its trace, and therefore also the $E_k$, are invariant.
We explicitly consider only the case when there is a finite number $g$ of non-degenerate forbidden gaps $(E_2,E_3)$, $(E_4,E_5)$, \dots, so that $E_{2k}=E_{2k+1}$ for all $k > g$.

\begin{figure}
    \includegraphics[width=0.4\textwidth]{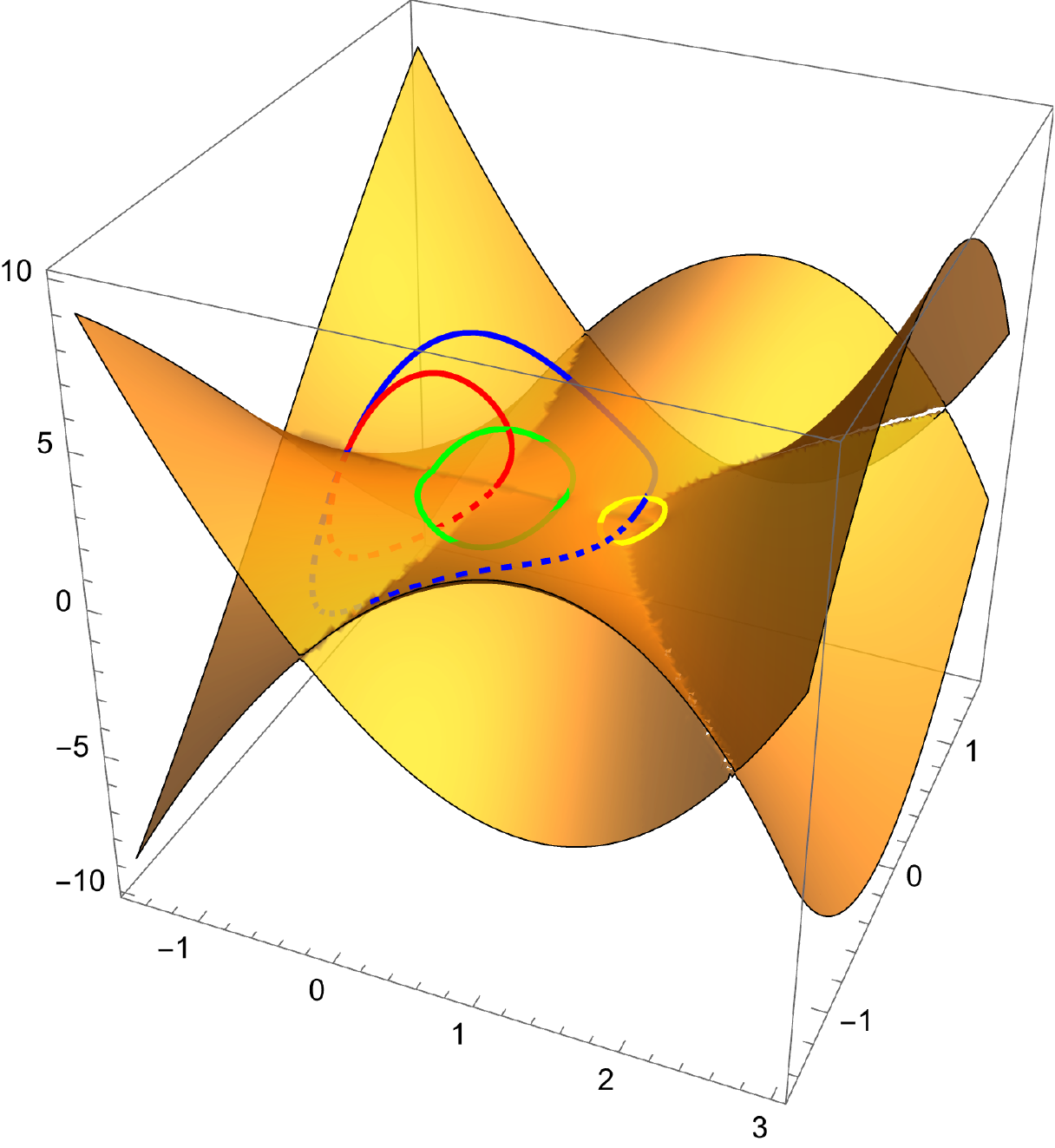}
    \caption{Riemann surface of genus 2 for the initial condition $u_0(x)=\sin{x}$, truncated to a 2-gap solution. The horizontal axes show the real and imaginary parts of $\lambda$ and the vertical axis shows the real part of $\mu$, as defined by (\ref{eq:riemannsurface}).}
    \label{fig:riemannsurface}
\end{figure}

\begin{figure}
\vspace{1em}
    \centering
    \tikzset{every picture/.style={line width=0.45pt}} %set default line width to 0.75pt        
    \begin{tikzpicture}[x=0.75pt,y=0.75pt,yscale=-0.8,xscale=0.8]
%uncomment if require: \path (0,245); %set diagram left start at 0, and has height of 245

    %Shape: Arc [id:dp2679343473434128] 
    \draw  [draw opacity=0][dash pattern={on 4.5pt off 4.5pt}] (193.14,130.54) .. controls (190.46,153.25) and (168.35,170.3) .. (142.35,169.1) .. controls (115.29,167.85) and (94.25,147.31) .. (95.36,123.22) -- (144.36,125.48) -- cycle ; \draw  [color={rgb, 255:red, 208; green, 2; blue, 27 }  ,draw opacity=1 ][dash pattern={on 4.5pt off 4.5pt}] (193.14,130.54) .. controls (190.46,153.25) and (168.35,170.3) .. (142.35,169.1) .. controls (115.29,167.85) and (94.25,147.31) .. (95.36,123.22) ;  
    %Shape: Arc [id:dp48995982474070776] 
    \draw  [draw opacity=0][dash pattern={on 4.5pt off 4.5pt}] (381.92,125.36) .. controls (381.97,126.24) and (382,127.12) .. (382,128) .. controls (382,173.84) and (314.26,211) .. (230.7,211) .. controls (147.14,211) and (79.4,173.84) .. (79.4,128) -- (230.7,128) -- cycle ; \draw  [color={rgb, 255:red, 9; green, 23; blue, 235 }  ,draw opacity=1 ][dash pattern={on 4.5pt off 4.5pt}] (381.92,125.36) .. controls (381.97,126.24) and (382,127.12) .. (382,128) .. controls (382,173.84) and (314.26,211) .. (230.7,211) .. controls (147.14,211) and (79.4,173.84) .. (79.4,128) ;  
    %Shape: Arc [id:dp507635525157305] 
    \draw  [draw opacity=0] (79.4,128) .. controls (79.34,127.12) and (79.31,126.24) .. (79.3,125.36) .. controls (78.91,79.52) and (146.33,41.78) .. (229.89,41.07) .. controls (313.44,40.36) and (381.5,76.94) .. (381.89,122.77) -- (230.6,124.07) -- cycle ; \draw  [color={rgb, 255:red, 9; green, 23; blue, 235 }  ,draw opacity=1 ] (79.4,128) .. controls (79.34,127.12) and (79.31,126.24) .. (79.3,125.36) .. controls (78.91,79.52) and (146.33,41.78) .. (229.89,41.07) .. controls (313.44,40.36) and (381.5,76.94) .. (381.89,122.77) ;  
    \draw  [color={rgb, 255:red, 24; green, 36; blue, 230 }  ,draw opacity=1 ] (186.78,50.4) -- (173.09,47.88) -- (183.33,38.44) ;
    %Shape: Ellipse [id:dp36404344803171473] 
    \draw  [color={rgb, 255:red, 200; green, 190; blue, 28 }  ,draw opacity=1 ] (345.2,129) .. controls (345.2,108.9) and (377.57,92.6) .. (417.5,92.6) .. controls (457.43,92.6) and (489.8,108.9) .. (489.8,129) .. controls (489.8,149.1) and (457.43,165.4) .. (417.5,165.4) .. controls (377.57,165.4) and (345.2,149.1) .. (345.2,129) -- cycle ;
    \draw  [color={rgb, 255:red, 200; green, 190; blue, 28 }  ,draw opacity=1 ] (419.37,86.64) -- (431.28,93.85) -- (418.37,99.05) ;
    %Shape: Ellipse [id:dp2062201995547479] 
    \draw  [color={rgb, 255:red, 65; green, 117; blue, 5 }  ,draw opacity=1 ] (158.6,133) .. controls (158.6,107.15) and (191.19,86.2) .. (231.4,86.2) .. controls (271.61,86.2) and (304.2,107.15) .. (304.2,133) .. controls (304.2,158.85) and (271.61,179.8) .. (231.4,179.8) .. controls (191.19,179.8) and (158.6,158.85) .. (158.6,133) -- cycle ;
    \draw  [color={rgb, 255:red, 65; green, 117; blue, 5 }  ,draw opacity=1 ] (242.97,81.44) -- (254.88,88.65) -- (241.97,93.85) ;
    %Shape: Arc [id:dp8783741088641372] 
    \draw  [draw opacity=0] (95.36,127.66) .. controls (96.53,104.83) and (117.45,86.35) .. (143.47,85.82) .. controls (169.83,85.28) and (191.71,103.35) .. (193.32,126.54) -- (144.36,129.48) -- cycle ; \draw  [color={rgb, 255:red, 208; green, 2; blue, 27 }  ,draw opacity=1 ] (95.36,127.66) .. controls (96.53,104.83) and (117.45,86.35) .. (143.47,85.82) .. controls (169.83,85.28) and (191.71,103.35) .. (193.32,126.54) ;  
    \draw  [color={rgb, 255:red, 208; green, 2; blue, 27 }  ,draw opacity=1 ] (118.99,100.23) -- (105.28,102.66) -- (111.56,90.24) ;
    
    % Text Node
    \draw (108,122.4) node [anchor=north west][inner sep=0.75pt]    {$E_{1}$};
    % Text Node
    \draw (164,123.4) node [anchor=north west][inner sep=0.75pt]    {$E_{2}$};
    % Text Node
    \draw (276,122.4) node [anchor=north west][inner sep=0.75pt]    {$E_{3}$};
    % Text Node
    \draw (449.6,119.4) node [anchor=north west][inner sep=0.75pt]    {$E_{5}$};
    % Text Node
    \draw (352.4,120.2) node [anchor=north west][inner sep=0.75pt]    {$E_{4}$};

    % Text Node
    \draw [color={rgb, 255:red, 65; green, 117; blue, 5 }  ,draw opacity=1 ] (200,75) node [anchor=north west][inner sep=0.75pt]    {$a_1$};
    % Text Node
    \draw [color={rgb, 255:red, 200; green, 190; blue, 28 }  ,draw opacity=1 ] (400,75) node [anchor=north west][inner sep=0.75pt]    {$a_2$};
    % Text Node
    \draw [color={rgb, 255:red, 208; green, 2; blue, 27 }  ,draw opacity=1 ] (120,70) node [anchor=north west][inner sep=0.75pt]    {$b_1$};
    % Text Node
    \draw [color={rgb, 255:red, 9; green, 23; blue, 235 },draw opacity=1 ] (108,50) node [anchor=north west][inner sep=0.75pt]    {$b_2$};

\end{tikzpicture}

    \caption{Schematic of the basis of contours for a $g=2$ Riemann surface, where dashed lines show the parts of the contour taken on the second sheet. The colours are as for figure \ref{fig:riemannsurface}.}
    \label{fig:contours}
\end{figure}
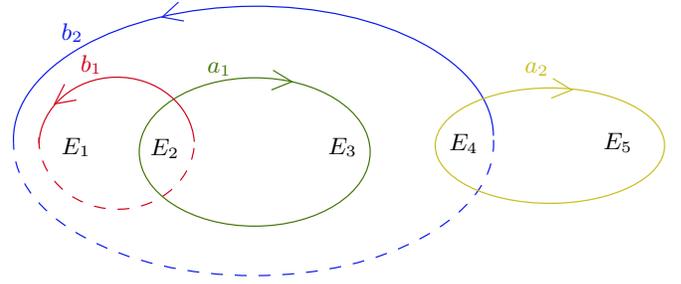

The hyperelliptic curve
\begin{equation}
\label{eq:riemannsurface}
    \mu^2 = \prod_{j=1}^{2g+1}(\lambda-E_j)
\end{equation}
defines a Riemann surface of two sheets, with a branch point at each $E_j$ (see figure \ref{fig:riemannsurface}). The \emph{genus} of this surface is simply the number of gaps $g$. It is then possible to define a basis of contours $a_j$ and $b_j$ ($1\leq j \leq g$) for the Riemann surface such that any contour can be expressed, up to continuous deformations, as a sum of the $a_j$ and $b_j$. Such a choice of basis is not unique, and will have implications for the final results -- see discussions of the wave basis and soliton basis in \citet{osborne2010nonlinear}. Our convention is shown in figure \ref{fig:contours}. We also define a basis of holomorphic differentials on this surface
\begin{equation}
    \mathring{\omega}_k = \frac{\lambda^{g-k}\mathrm{d}\lambda}{\mu}
\end{equation}
and then make a linear transformation to the canonical basis $\omega_k=\sum_j \mathbf{c}_{kj}\mathring{\omega}_j$ such that
\begin{equation}
\int_{a_j}\omega_k = 2\pi i \delta_{jk},\quad{j,k=1,\dots,g}.
\end{equation}
In this new basis, we define the \emph{period matrix} of the Riemann surface
\begin{equation}
    \mathbf{B}_{jk} = \int_{b_j}\omega_k.
\end{equation}
It can be shown that this matrix is symmetric $\mathbf{B}_{jk}=\mathbf{B}_{kj}$ with all entries having strictly negative real part. It is then the case that the transformed variable $\vartheta$ can be written as
\begin{equation}
\label{eq:theta}
    \vartheta(x,t) = \theta(\bm{U}x+\bm{W}t+\bm{D},\mathbf{B})
\end{equation}
where we define the Riemann theta function
\begin{equation}
\label{eq:riemanntheta}
\theta(\bm{z},\mathbf{B})=\sum_{\bm{m}\in\mathbb{Z}^g}\exp{\left(\frac{1}{2}\bm{m}^T\mathbf{B}\bm{m}+\bm{z}^T\bm{m}\right)}.
\end{equation}
For $\mathbf{B}$ real and $\bm{z}$ imaginary, this gives real values by symmetry.
The wavenumber vector $\bm{U}$ and frequency vector $\bm{W}$ are calculated as
\begin{align}
    \bm{U}_j &= 2i\mathbf{c}_{j1},\\
    \bm{W}_j &= -2i\left(\mathbf{c}_{j2}+\frac{1}{2}\mathbf{c}_{j1}\sum_{k=1}^{2g+1} E_k\right).
\end{align}
By construction, the $\bm{U}_j$ must be integers, but the frequencies $\bm{W}_j$ will be incommensurate in general.
Both $\bm{U}$ and $\bm{W}$ are purely imaginary.
Since they are calculated only from the $E_k$, all of $\mathbf{B}$, $\bm{U}$ and $\bm{W}$ are constant as the system evolves.
The value of the vector of phases $\bm{D}$, conversely, depends on the particular state at time $t=0$ (and its evolution is absorbed into $\bm{W}$). 

To find the phases $\bm{D}$, it is necessary to define a second set of eigenvalues for the Sturm-Liouville problem (\ref{eq:eigenvalueproblem}), now with the boundary conditions $\psi(0)=\psi(2\pi)=0$. This discrete set of eigenvalues $\lambda_1, \lambda_2, \dots$ lies in the gaps so that $E_2<\lambda_1<E_3$, $E_4<\lambda_2<E_5$ etc. These eigenvalues are not constant as the state evolves, and depend on the time of measurement.
The formula for $\bm{D}$ is then given by \citep{belokolos1994algebro}
\begin{equation}
    \bm{D}_j=-\sum_{k=1}^g\int_\infty^{\lambda_k}\omega_j-\sum_{k=1}^g \mathbf{B}_{jk}+i\pi j,
\end{equation}
where here $\lambda_k$ represents a point on the surface with $\lambda=\lambda_k$, with care be taken to evaluate the integral on the correct sheet. $\bm{D}$ is also purely imaginary.

Finally, this allows us to define a Koopman eigenfunction for  (\ref{eq:kdv}) for each $\bm{m}\in\mathbb{Z}^g$:
\begin{equation}
    \label{eq:eigenfunction}
    \varphi_{\bm{m}}(u_0) := \exp{\left(\bm{D}^T\bm{m}\right)}
\end{equation}
which then evolves as
\begin{equation*}
    \varphi_{\bm{m}}(u(\cdot,t) )= \exp{\left(\bm{W}^T\bm{m}\,t\right)}\exp{\left(\bm{D}^T\bm{m}\right)}
\end{equation*}
and thus has Koopman eigenvalue $\bm{W}^T\bm{m}$, an integer linear combination of the fundamental frequencies. As the frequencies are incommensurate, for genus $g=2$ and greater, these eigenvalues densely fill the imaginary axis. This is a significant complication over previously studied PDEs. Note that $\varphi_{\bm{m}_1}(u)\,\varphi_{\bm{m}_2}(u)=\varphi_{\bm{m}_1+\bm{m}_2}(u)$, and $\varphi_{\bm{0}}(u)=1$.

\section{Koopman decompositions}
\label{sec:results}
Clearly the expression (\ref{eq:theta}) is directly a convergent Koopman decomposition for $\vartheta(x,t)$, as it is a sum of terms whose time dependence is purely exponential. We can write it as
\begin{equation}
    \label{eq:thetaform}
\vartheta(x,0)=\sum_{\mathbf{m}\in\mathbb{Z}^g}e^{\bm{U}^T\bm{m}x}\exp{\left(\frac{1}{2}\bm{m}^T\mathbf{B}\bm{m}\right)}\varphi_\mathbf{m}(u_0).
\end{equation}
It is somewhat more involved to obtain a decomposition of $u$, but this is still possible so long as $\vartheta$ is sufficiently small, via (\ref{eq:hirota}):

\begin{widetext}
\begin{align}
\label{eq:koopmandecomposition}
\begin{split}
    u_0(x)
    &=2\partial_x^2 \log{\left[1+\sum_{\bm{m}\in\mathbb{Z}^g\backslash\{\bm{0}\}}\exp{\left(\frac{1}{2}\bm{m}^T\mathbf{B}\bm{m}+\bm{U}^T\bm{m}x\right)}\varphi_{\bm{m}}(u_0)\right]}\\
    &=-2\partial_x^2\left[\sum_{n=1}^{\infty}\frac{(-1)^{n}}{n}\sum_{\bm{m_1},\dots,\bm{m}_n\in\mathbb{Z}^g\backslash\{\bm{0}\}}\exp{\left\{\sum_{q=1}^n\left(\frac{1}{2}\bm{m}_q^T\mathbf{B}\bm{m}_q+\bm{U}^T\bm{m}_q x\right)\right\}}\prod_{q=1}^n\varphi_{\bm{m}_q}(u_0)\right]\\
    &=\sum_{\bm{m}\in\mathbb{Z}^g}\left(-2   \left(\bm{U}^T\bm{m}\right)^2  e^{\bm{U}^T\bm{m} x} \sum_{n=1}^{\infty}\frac{(-1)^{n}}{n}     \mathlarger{\sum}_{\substack{\bm{m_1},\dots,\bm{m}_n\in\mathbb{Z}^g\backslash\{\bm{0}\}\\ \sum_q \bm{m}_q = \bm{m}}} \exp{\left\{\sum_{q=1}^n\frac{1}{2}\bm{m}_q^T\mathbf{B}\bm{m}_q\right\}}\right)\varphi_{\bm{m}}(u_0).
\end{split}
\end{align}
\end{widetext}
%This complex formula bears a striking resemblance to that derived by \citet{balabane2021koopman} for the Burgers equation, in that we must sum over products of the Koopman eigenfunctions derived in section \ref{sec:eigenfunctions}.
This complicated series is absolutely convergent when $0<\vartheta(x,t)<2$. For larger $\vartheta$, other expansions could be found, using only the Koopman eigenfunctions given in the previous section.
To summarise this expression, we have found a Koopman decomposition for $u$ using Koopman eigenfunctions $\varphi_{\bm{m}}(u_0)$ with Koopman eigenvalues $\bm{W}^T\bm{m}$.
The corresponding Koopman modes are
\begin{widetext}
\begin{equation}
    -2   \left(\bm{U}^T\bm{m}\right)^2  e^{\bm{U}^T\bm{m} x} \sum_{n=1}^{\infty}\frac{(-1)^{n}}{n}     \mathlarger{\sum}_{\substack{\bm{m_1},\dots,\bm{m}_n\in\mathbb{Z}^g\backslash\{\bm{0}\}\\ \sum_q \bm{m}_q = \bm{m}}}    \exp{\left\{\sum_{q=1}^n\frac{1}{2}\bm{m}_q^T\mathbf{B}\bm{m}_q\right\}}
\end{equation}
\end{widetext}
Notice that these are purely sinusoidal in $x$. The Koopman modes depend on $\bm{U}$ and $\mathbf{B}$, which are functions of the Riemann surface and therefore of the $g$-torus to which the dynamics are constrained in phase space, but the Koopman modes do not depend on the choice of initial conditions beyond this.

As a concrete example, we consider the initial condition $u_0(x) = \sin{x}$.
Despite the simplicity of this choice, it gives an apparently infinite number of non-degenerate gaps (see figure \ref{fig:monodromy}), but using only $g=2$ or $g=3$ results in good agreement. With $g=4$, not shown here, the reconstructed solution is virtually indistinguishable from the initial condition. 
Only two solitons are visible per spatial period in a numerical solution; the genus of the Riemann surface is not the number of solitons.
For $g=2$ we find numerically that
\begin{equation}
    \bm{U}=(-1i,-2i),\quad\bm{W}\approx(-0.036,1.930i),
    \end{equation}
    \begin{equation}\mathbf{B}\approx
    \begin{pmatrix}
    -3.171&-1.930\\
    -1.930&-7.247
    \end{pmatrix},
\end{equation}
and for $g=3$
\begin{equation}
    \bm{U}=(-1i,-2i,-3i),\quad\bm{W}\approx(-0.036i,1.931i,6.741i),
\end{equation}
\begin{equation}
\label{eq:Bans}
    \mathbf{B}\approx
    \begin{pmatrix}
    -3.171&-1.929&-1.321\\
    -1.929&-7.247&-3.190\\
    -1.321&-3.190&-12.810
    \end{pmatrix}.
\end{equation}

\begin{figure}
    \centering
    \includegraphics[width=\columnwidth]{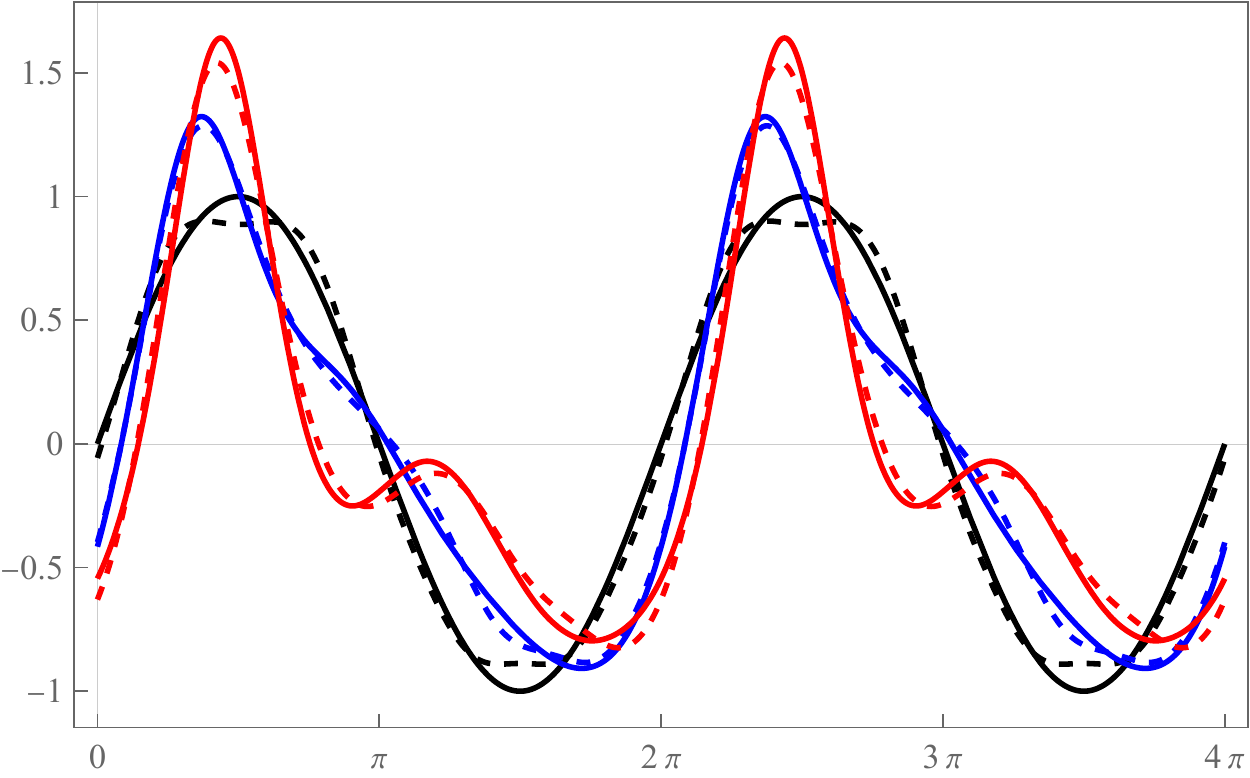}
    \caption{Numerical solution $u$ (solid) and Koopman reconstruction (dashed) at $t=0$ (black), $t=0.5$ (blue) and $t=1$ (red) using $g=2$, i.e. assuming the dynamics are constrained to a 2 dimensional quasiperiodic invariant torus.
    The Koopman decomposition (\ref{eq:koopmandecomposition}) is truncated after $n=4$ with $\bm{m},\bm{m}_q\in \left\{ -3,...,3 \right\}^2\backslash\bm{0}$.}
    \label{fig:reconstructed2}
\end{figure}

\begin{figure}
    \centering
    \includegraphics[width=\columnwidth]{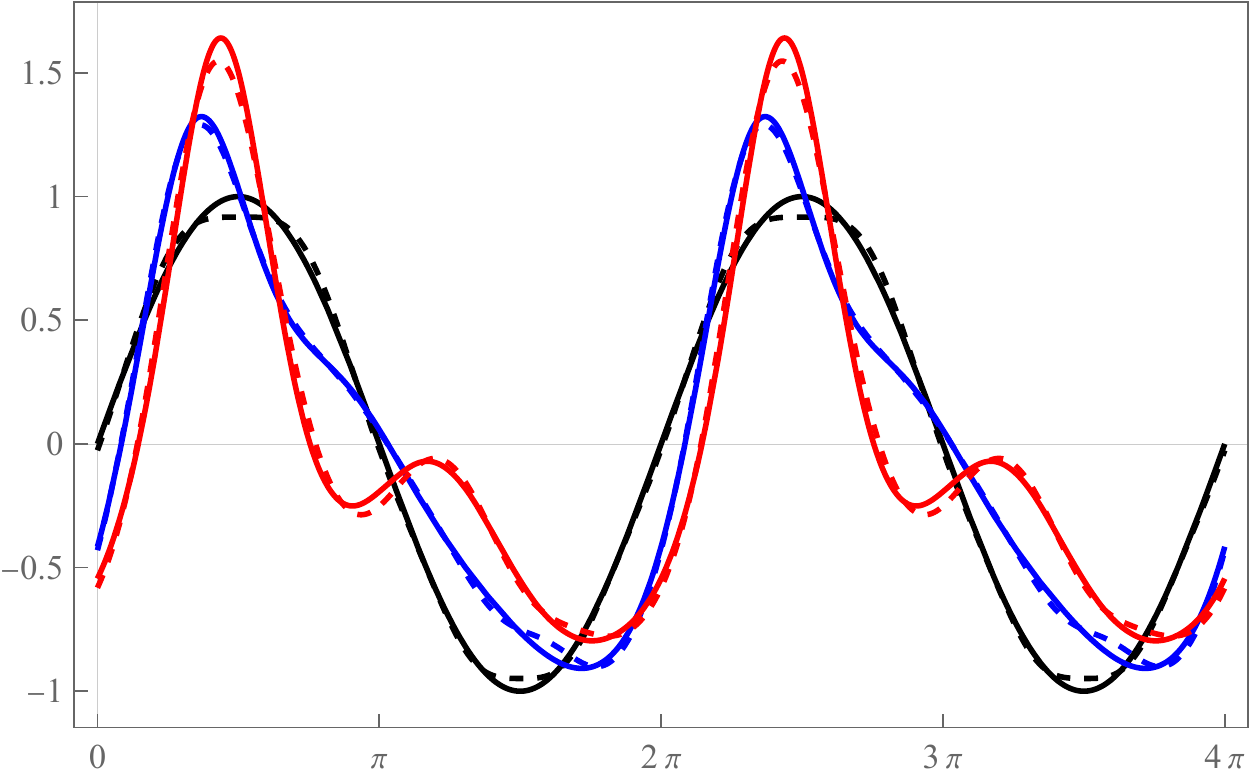}
    \caption{As for figure \ref{fig:reconstructed2} but with $g=3$, and $\bm{m},\bm{m}_q\in \left\{ -3,...,3 \right\}^3\backslash\bm{0}$.}
    \label{fig:reconstructed3}
\end{figure}

The reconstructed $u(x,t)$ given by a finite truncation of the Koopman decomposition (\ref{eq:koopmandecomposition}) is shown in figures \ref{fig:reconstructed2} and \ref{fig:reconstructed3}.
Including terms higher than $n=4$ in the series may increase the accuracy of these, but the Koopman modes become prohibitively expensive to evaluate numerically.

\section{Dynamic mode decomposition}
\label{sec:dmd}
Given a discrete time-series of snapshots $\{\vv_j\}_j$ from some dynamical system, DMD seeks to find a linear map $A$ such that $\vv_{j + 1} = A \vv_{j}$. In practice, DMD finds an eigendecomposition of $A$ in which each mode of the decomposition has an associated amplitude $a_k$, spatial pattern $\uu_k$ and growth rate $\lambda_k$. As in the case of the Koopman eigenvalues, we expect $\lambda_k= i \omega_k$ to be purely imaginary because the system is Hamiltonian. We can use our expectation of purely imaginary eigenvalues as a heuristic for a well-resolved mode: if $|Re(\lambda_k))| \gg 0$ we infer that $\omega_k$ is an inaccurate guess.
Then if the time between snapshots is $\tau$, we can reconstruct the evolution of the system as
\[ \vv_j = \sum_k e^{i \omega_k j \tau} \uu_k. \]

In the original and most basic form of the algorithm, the number of DMD modes that come from the spectral decomposition will be equal to the spatial dimension. We can increase both the robustness and number of discovered of DMD modes found with delay embedding, a higher-order extension in which temporal resolution is substituted for spatial resolution \cite{schmid_dynamic_2022}. We found much better results when employing delay embedding; we used $20$ delays.

As DMD is designed to detect the important temporal frequencies of the dynamics, it should be possible to use it to reconstruct an approximation $\Tilde{\bm{W}}$ for $\bm{W}$ from a time-series. However, as discussed in section \ref{sec:eigenfunctions}, the Koopman eigenvalues densely fill the imaginary axis, and so the results of DMD are obscured.
For example, if we expect the solution to be well-represented by an invariant 2-torus --- and the DMD eigenfrequencies $\omega_k$ are sufficiently well-resolved --- we expect to see  $i\omega_k = n_{1k} \bm{W}_1 + n_{2k} \bm{W}_2$, for many different $n_{1k},n_{2k}\in \mathbb{Z}$.
However, as hinted at in section \ref{sec:results}, the amplitudes associated with low $n_{jk}$ should be larger.
Aided by knowing the relative amplitude of each DMD mode (to empirically identify ``important'' modes), we can guess the smaller of the $\Tilde{\bm{W}}_j$ to be the gap between eigenvalues (and the smaller high amplitude mode) and the larger $\Tilde{\bm{W}}_j$ to be the second largest high amplitude mode. We can extend this argument in the obvious way for $g > 2$.

\begin{figure}
    \centering
    \includegraphics[width=\columnwidth]{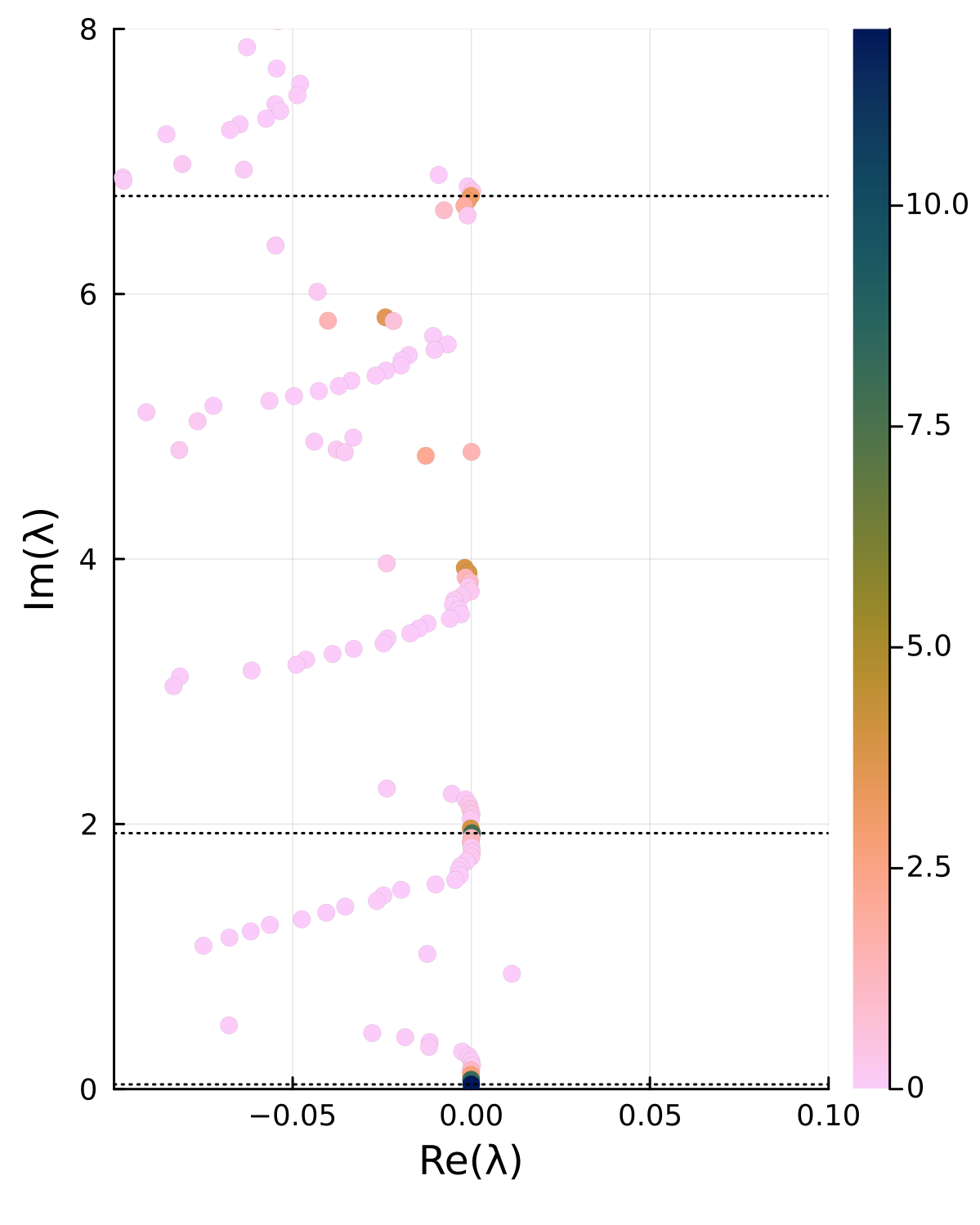}
    \caption{DMD eigenvalues $\lambda$ from analysis of a timeseries of $u$, where points are coloured to represent the amplitude ($|a|$) of each DMD mode (the colorbar saturates at 10). We plot only eigenvalues with real part less than 0.1 in magnitude, which excludes a number of spurious modes. Horizontal lines at $\Tilde{\bm{W}}_1 = 0.03558i$,  $\Tilde{\bm{W}}_2 = 1.9312i$, and $\Tilde{\bm{W}}_3 = 6.7394i$ mark the DMD approximations of $\bm{W}$. Prominent eigenvalues are also evident near $2\mathbf{W}_1$, $2\mathbf{W}_2$, $\mathbf{W}_1+\mathbf{W}_2$, $\mathbf{W}_3-\mathbf{W}_2$, etc. }
    \label{fig:eigsu}
\end{figure}

We apply DMD to a numerical solution of (\ref{eq:kdv}) with initial condition $u_0(x) = \sin{x}$ as in section \ref{sec:results}. Our numerical simulation has a length of 450 time units, with a time resolution $\tau = 0.1$. We find very good agreement in the identification of $\bm{W}$ as computed analytically to those found in DMD:
\begin{equation}
\label{eq:dmdfrequencies}
    \Tilde{\bm{W}} \approx (0.036i,1.931i, 6.739i).
\end{equation}
Note that the differing signs represent a degeneracy of the formulation, these could be recovered in the analytic method by using a different basis of integration contours.
Figure \ref{fig:eigsu} shows the DMD eigenvalues as well as their relative amplitudes.

Additionally, given that we know the Riemann theta function form 
(\ref{eq:riemanntheta}), we can exploit the Hirota transform (\ref{eq:hirota}). By applying DMD to a time-series of $\vartheta$ rather than $u$, we can recover all parameters for the Riemann theta function, i.e. $\bm{U}$, $\bm{W}$ and $\mathbf{B}$. The details of this procedure are given in the appendix. We find
\begin{equation}
    \Tilde{\bm{B}} \approx \begin{pmatrix}
        -3.077 & -1.907 \\
        -1.907 & -7.150
    \end{pmatrix}
\end{equation}
which is approximately consistent with the values computed analytically, given in (\ref{eq:Bans}).

\begin{figure*}
    \centering
    \includegraphics[width=0.44\textwidth]{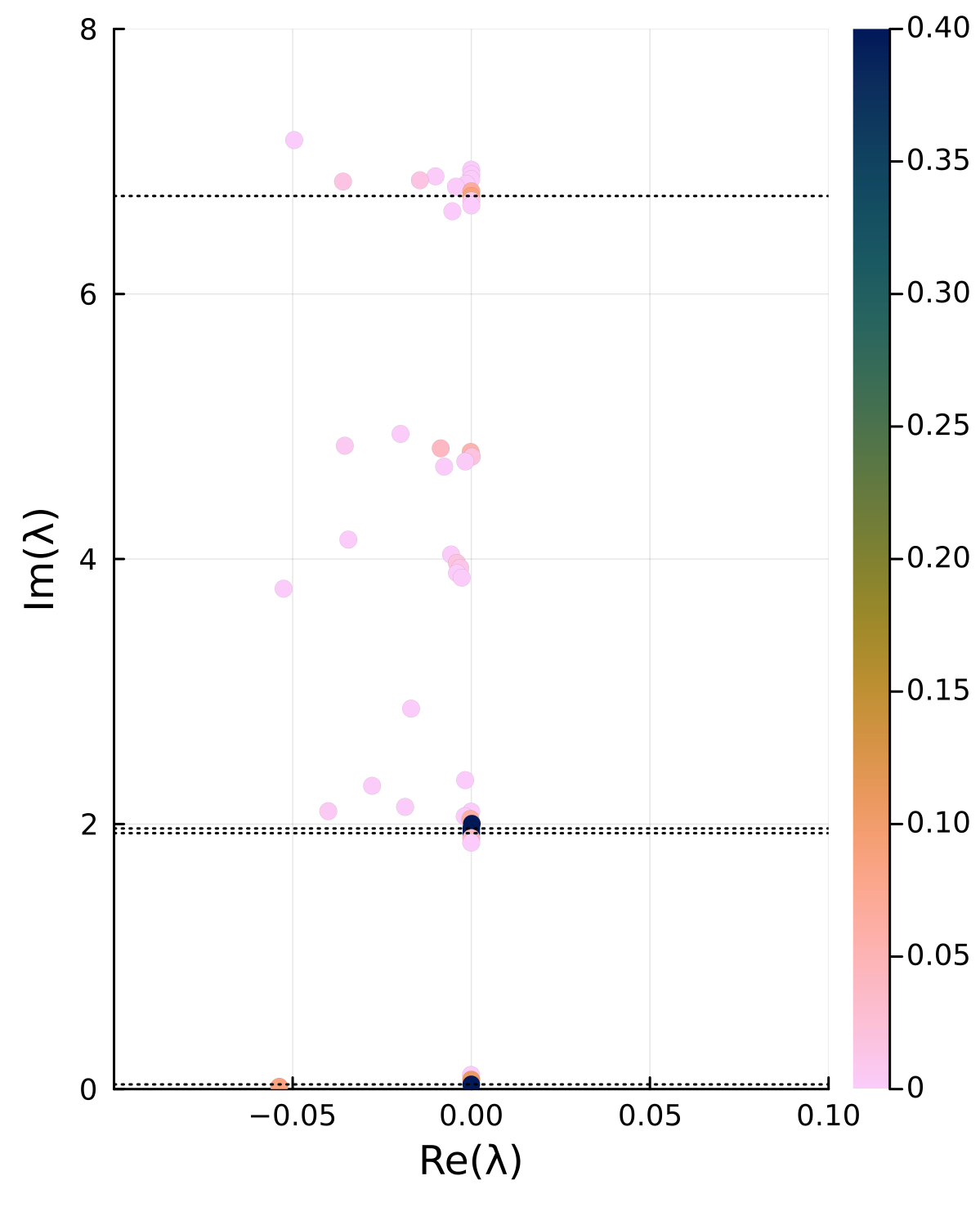}
    \includegraphics[width=0.55\textwidth]{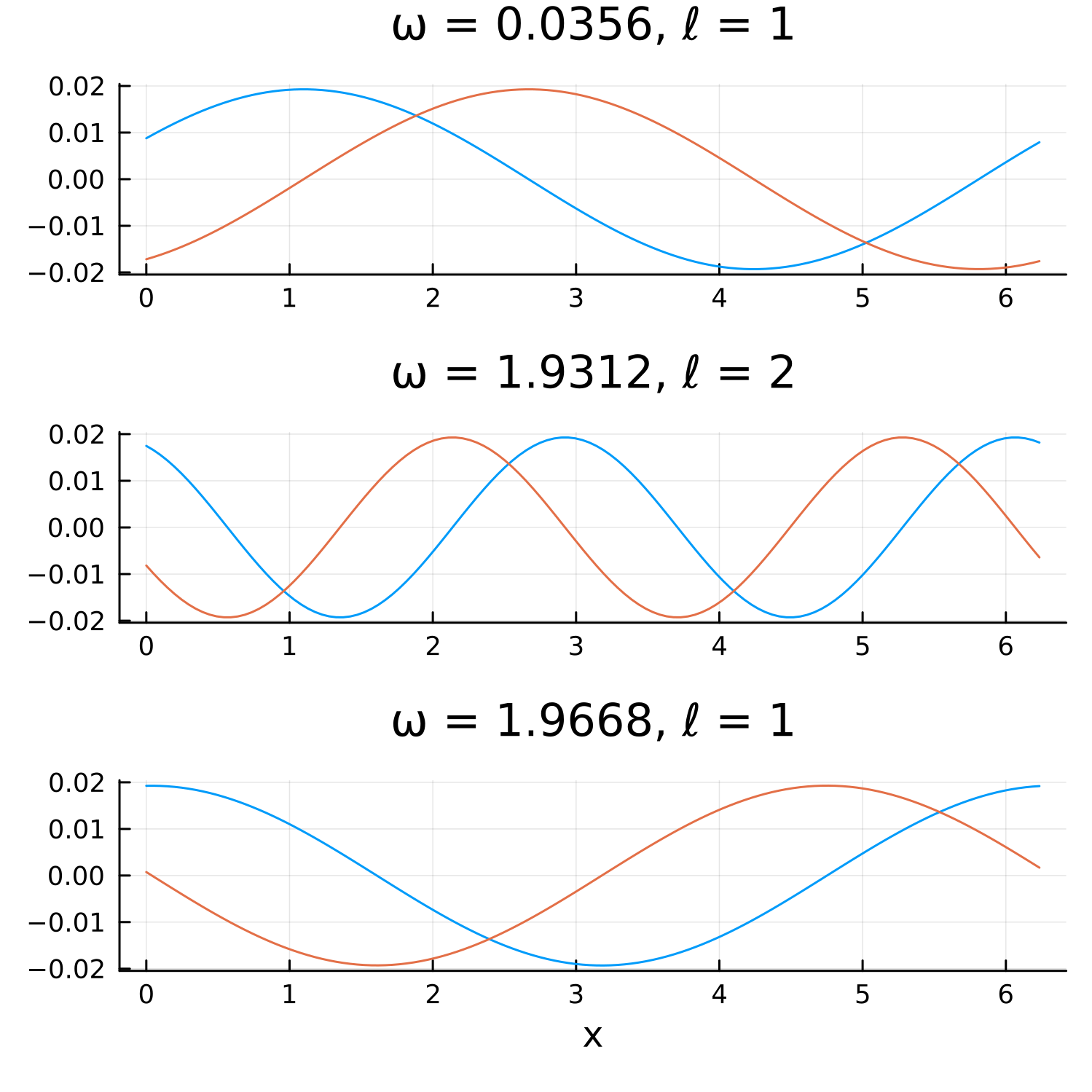}
    \caption{Left: As for \ref{fig:eigsu} except that data analyzed was of $\vartheta$ rather than $u$. Horizontal lines show $\omega_j$ for the well-resolved modes: $\Tilde{\bm{W}}_1 = 0.0356i$, $\Tilde{\bm{W}}_2 = 1.9312i$, $\Tilde{\bm{W}}_1 + \Tilde{\bm{W}}_2 = 1.9668i$, and $\Tilde{\bm{W}}_3 = 6.7394i$. Right: DMD spatial modes corresponding to the given frequencies. The real part of the mode is shown in blue, the imaginary part in orange. For each pattern, we also compute the wavenumber $\ell$.}
    \label{fig:eigstheta}
\end{figure*}

\section{Discussion}
\label{sec:discussion}

We have performed a Koopman decomposition of the periodic KdV equation. This is almost immediate once the convoluted but well-defined process of periodic inverse scattering is performed. Additionally, we have shown how this result relates to the output of DMD for such a system. DMD gives a very large number of near-imaginary eigenvalues associated with the different harmonics of the nonlinearly interacting waves, which correspond to the Koopman eigenvalues found analytically, which densely fill the imaginary axis.

Further, by exploiting the $\theta$-function representation of the solution, we are able to use DMD to approximately recover the necessary parameters.

We note in passing that since we expect purely imaginary eigenvalues in our system, it is a potential use-case for the physics-informed DMD method \citep{baddoo2021physics} of finding a unitary matrix $A$ to fit the data. However, we found that this obfuscates the results, as it prevents the use of the real part of the eigenvalue as a measure for how well-resolved a given mode is.

The analytic results of this paper could be extended to other integrable PDEs which admit Lax pairs, such as the nonlinear Schr\"odinger equation, the sine-Gordon equation or the Kadomtsev-Petviashvili equation. The latter could be particularly insightful since it describes two-dimensional wave fields, a  significant increase in complexity over the one-dimensional PDEs which have been studied heretofore.

\section*{Acknowledgements}
This work started life at the Geophysical Fluid Dynamics summer school at Woods Hole Oceanographic Institution, and section \ref{sec:dmd} represents a small part of the fellow's project of CV. The authors wish to thank Peter Schmid for his help with this project, and everyone at GFD for many fruitful discussions. JPP would like to thank Al Osborne for some useful pointers.

\section*{Author Declarations}
The authors have no conflicts to disclose.

\section*{Data Availability}
The data that supports the findings of this study are available within the supplementary material. 

\section*{Appendix}
Here we give a brief overview of the procedure to recover  the parameters for the Riemann theta function, i.e. $\bm{U}$, $\bm{W}$ and $\mathbf{B}$, from a time-series of $\vartheta$.

Again assuming $g=2$, we determine $\bm{W}_j$ from $\vartheta$ as we did with $u$. The wavenumber vector $\bm{U}_j$ is recovered using the same idea; as each DMD frequency $\omega_k$ is an integer linear combination of $i\bm{W}_j$, we expect that each DMD spatial mode $\uu_k$ will be a pure sinusoid with wavenumber $\ell_j$ satisfying $i\ell_j = n_{1k} \bm{U}_1 + n_{2k} \bm{U}_2$. To solve for entries of the period matrix $\mathbf{B}$, we will need 3 DMD modes, where we can infer $n_{1k}, n_{2k}$ for each mode $k$. We construct an invertible matrix $\bm{M}$ where each row $\bm{M}_k$ has entries $(n_{1k}^2, 2 n_{1k} n_{2k},n_{2k}^2 )$. Then letting $\mathbf{c}$ be a vector such that $c_k = 2\log(a_k)$, we solve
\begin{equation}
    \bm{M} \mathbf{b} = \mathbf{c}, \: \mathbf{b} = (\mathbf{B}_{11}, \mathbf{B}_{21}, \mathbf{B}_{22})
\end{equation} 
for $\mathbf{B}$.
We note that for $g > 2$, we can still determine all parameters of the Riemann theta function give enough well-resolved DMD nodes. However, given that the symmetric $g\times g$ matrix $\mathbf{B}$ will have $(g + 1)g/2$ unique entries, we will need to identify the same number of well-resolved DMD modes which can be a nontrivial task even for small $g$.

We apply our DMD analysis to KdV data with initial condition $u_0(x) = \sin{x}$, where we analyze the value of $\vartheta$, rather than $u$. 
To three decimal places, we recover the same values for the frequencies (\ref{eq:dmdfrequencies}).
Figure \ref{fig:eigstheta} shows the eigenvalues, along with the three modes corresponding to frequencies $\tilde{\bm{W}}_1$ ,$\tilde{\bm{W}}_2$, and $\tilde{\bm{W}}_1+\tilde{\bm{W}}_2$ which were used to determine $\tilde{\mathbf{B}}$.
The DMD spectrum for $\vartheta$ is much cleaner than for $u$, which shows that the Hirota transform has in some sense simplified the dynamics.

% compute similar values for $\Tilde{\bm{W}}$:
%\begin{equation}
%    \Tilde{\bm{W}}_1 \approx 0.0356, \: \Tilde{\bm{W}}_2 \approx %1.9312, \: \Tilde{\bm{W}}_3 \approx 6.7394
%\end{equation}
% In this case, the value for $\Tilde{\bm{W}}_3$ was less obvious to see from DMD data and amplitude comparison (see figure \ref{fig:eigstheta}, left) as a value $\omega \approx 4.837$ also was a relatively high amplitude mode. However, as we know that $\Tilde{\bm{W}}_2 \approx 1.7938$, we can check that $4.837$ is not one of the fundamental frequencies as $4.837 \approx \Tilde{\bm{W}}_3 - \Tilde{\bm{W}}_2$. 

%We compute the approximate coefficients of $\bm{B}$, using the three DMD modes with frequencies $\omega_1 = 0.0356 \approx \bm{W}_1$, $\omega_2 = 1.9312 \approx \bm{W}_2$, and $\omega_3 = 1.9668 \approx \bm{W}_2 + \bm{W}_1$. We plot their spatial modes (where we also infer the values of $\bm{C}$) in figure \ref{fig:eigstheta} (right). 

\bibliographystyle{unsrtnat}
\bibliography{references}

\begin{thebibliography}{30}
\providecommand{\natexlab}[1]{#1}
\providecommand{\url}[1]{\texttt{#1}}
\expandafter\ifx\csname urlstyle\endcsname\relax
  \providecommand{\doi}[1]{doi: #1}\else
  \providecommand{\doi}{doi: \begingroup \urlstyle{rm}\Url}\fi

\bibitem[Koopman(1931)]{koopman1931hamiltonian}
Bernard~O Koopman.
\newblock Hamiltonian systems and transformation in hilbert space.
\newblock \emph{Proceedings of the National Academy of Sciences}, 17\penalty0
  (5):\penalty0 315--318, 1931.

\bibitem[Mezi\'c(2005)]{Mezic2005}
I.~Mezi\'c.
\newblock Spectral properties of dynamical systems, model reduction and
  decompositions.
\newblock \emph{Nonlinear Dynam.}, 41:\penalty0 309--325, 2005.
\newblock \doi{10.1007/s11071-005-2824-x}.

\bibitem[Mezi\'c(2013)]{Mezic2013}
I.~Mezi\'c.
\newblock Analysis of fluid flows via spectral properties of the {Koopman}
  operator.
\newblock \emph{Ann. Rev. Fluid Mech.}, 45:\penalty0 357--378, 2013.
\newblock \doi{10.1146/annurev-fluid-011212-140652}.

\bibitem[Schmid(2010)]{Schmid2010}
P.~J. Schmid.
\newblock {Dynamic mode decomposition of numerical and experimental data}.
\newblock \emph{J. Fluid Mech.}, 656:\penalty0 5--28, 2010.
\newblock \doi{10.1017/S0022112010001217}.

\bibitem[Schmid et~al.(2011)Schmid, Li, Juniper, and
  Pust]{schmid2011applications}
Peter~J Schmid, Larry Li, Matthew~P Juniper, and O~Pust.
\newblock Applications of the dynamic mode decomposition.
\newblock \emph{Theoretical and Computational Fluid Dynamics}, 25\penalty0
  (1):\penalty0 249--259, 2011.

\bibitem[Kutz et~al.(2016)Kutz, Brunton, Brunton, and Proctor]{kutz2016dynamic}
J~Nathan Kutz, Steven~L Brunton, Bingni~W Brunton, and Joshua~L Proctor.
\newblock \emph{Dynamic mode decomposition: data-driven modeling of complex
  systems}.
\newblock SIAM, 2016.

\bibitem[Schmid(2022)]{schmid_dynamic_2022}
Peter~J. Schmid.
\newblock Dynamic {Mode} {Decomposition} and {Its} {Variants}.
\newblock \emph{Annual Review of Fluid Mechanics}, 54\penalty0 (1):\penalty0
  225--254, 2022.
\newblock \doi{10.1146/annurev-fluid-030121-015835}.
\newblock URL \url{https://doi.org/10.1146/annurev-fluid-030121-015835}.

\bibitem[Rowley et~al.(2009)Rowley, Mezi\'c, Bagheri, Schlatter, and
  Henningson]{Rowley2009}
C.~W. Rowley, I.~Mezi\'c, S.~Bagheri, P.~Schlatter, and D.~S. Henningson.
\newblock {Spectral analysis of nonlinear flows}.
\newblock \emph{J. Fluid Mech.}, 641:\penalty0 115--127, 2009.
\newblock \doi{10.1017/S0022112009992059}.

\bibitem[Nathan~Kutz et~al.(2018)Nathan~Kutz, Proctor, and
  Brunton]{kutz2018applied}
J~Nathan~Kutz, Joshua~L Proctor, and Steven~L Brunton.
\newblock Applied {Koopman} theory for partial differential equations and
  data-driven modeling of spatio-temporal systems.
\newblock \emph{Complexity}, 2018, 2018.

\bibitem[Page and Kerswell(2018)]{page2018koopman}
Jacob Page and Rich~R Kerswell.
\newblock Koopman analysis of {Burgers} equation.
\newblock \emph{Physical Review Fluids}, 3\penalty0 (7):\penalty0 071901, 2018.

\bibitem[Balabane et~al.(2021)Balabane, Mendez, and Najem]{balabane2021koopman}
Mikhael Balabane, Miguel~Alfonso Mendez, and Sara Najem.
\newblock Koopman operator for {Burgers's} equation.
\newblock \emph{Physical Review Fluids}, 6\penalty0 (6):\penalty0 064401, 2021.

\bibitem[Nakao and Mezi{\'c}(2020)]{nakao2020spectral}
Hiroya Nakao and Igor Mezi{\'c}.
\newblock Spectral analysis of the koopman operator for partial differential
  equations.
\newblock \emph{Chaos: An Interdisciplinary Journal of Nonlinear Science},
  30\penalty0 (11):\penalty0 113131, 2020.

\bibitem[Parker and Page(2020)]{parker2020koopman}
Jeremy~P Parker and Jacob Page.
\newblock Koopman analysis of isolated fronts and solitons.
\newblock \emph{SIAM Journal on Applied Dynamical Systems}, 19\penalty0
  (4):\penalty0 2803--2828, 2020.

\bibitem[Zabusky and Kruskal(1965)]{zabusky1965interaction}
Norman~J Zabusky and Martin~D Kruskal.
\newblock Interaction of ``solitons'' in a collisionless plasma and the
  recurrence of initial states.
\newblock \emph{Physical Review Letters}, 15\penalty0 (6):\penalty0 240, 1965.

\bibitem[Fermi et~al.(1955)Fermi, Pasta, Ulam, and Tsingou]{fermi1955studies}
Enrico Fermi, P~Pasta, Stanislaw Ulam, and Mary Tsingou.
\newblock Studies of the nonlinear problems.
\newblock Technical report, Los Alamos National Lab.(LANL), Los Alamos, NM
  (United States), 1955.

\bibitem[Lax(1976)]{lax1976almost}
Peter~D Lax.
\newblock Almost periodic solutions of the {KdV} equation.
\newblock \emph{SIAM review}, 18\penalty0 (3):\penalty0 351--375, 1976.

\bibitem[Belokolos et~al.(1994)Belokolos, Bobenko, Enolskii, Its, and
  Matveev]{belokolos1994algebro}
Eugene~D Belokolos, Alexander~I Bobenko, Viktor~Z Enolskii, Alexander~R Its,
  and Vladimir~B Matveev.
\newblock \emph{Algebro-geometric approach to nonlinear integrable equations},
  volume 550.
\newblock Springer, 1994.

\bibitem[Korteweg and {de Vries}(1895)]{korteweg1895xli}
Diederik~Johannes Korteweg and Gustav {de Vries}.
\newblock Xli. on the change of form of long waves advancing in a rectangular
  canal, and on a new type of long stationary waves.
\newblock \emph{The London, Edinburgh, and Dublin Philosophical Magazine and
  Journal of Science}, 39\penalty0 (240):\penalty0 422--443, 1895.

\bibitem[Gardner et~al.(1967)Gardner, Greene, Kruskal, and
  Miura]{gardner1967method}
Clifford~S Gardner, John~M Greene, Martin~D Kruskal, and Robert~M Miura.
\newblock Method for solving the korteweg-devries equation.
\newblock \emph{Physical Review Letters}, 19\penalty0 (19):\penalty0 1095,
  1967.

\bibitem[Benney(1966)]{benney1966}
D.~J. Benney.
\newblock Long non-linear waves in fluid flows.
\newblock \emph{Journal of Mathematics and Physics}, 45\penalty0
  (1-4):\penalty0 52--63, 1966.
\newblock \doi{https://doi.org/10.1002/sapm196645152}.
\newblock URL
  \url{https://onlinelibrary.wiley.com/doi/abs/10.1002/sapm196645152}.

\bibitem[Peregrine(1966)]{peregrine1966calculations}
D~Howell Peregrine.
\newblock Calculations of the development of an undular bore.
\newblock \emph{Journal of Fluid Mechanics}, 25\penalty0 (2):\penalty0
  321--330, 1966.

\bibitem[Karpman(1975)]{karpman1975non}
Vladimir~Iosifovich Karpman.
\newblock \emph{Non-linear waves in dispersive media: International series of
  monographs in natural philosophy}, volume~71.
\newblock Elsevier, 1975.

\bibitem[Kappeler and Topalov(2006)]{Kappeler2006}
T.~Kappeler and P.~Topalov.
\newblock {Global wellposedness of KdV in $H^{-1}({\mathbb T},{\mathbb R})$}.
\newblock \emph{Duke Mathematical Journal}, 135\penalty0 (2):\penalty0 327 --
  360, 2006.
\newblock \doi{10.1215/S0012-7094-06-13524-X}.
\newblock URL \url{https://doi.org/10.1215/S0012-7094-06-13524-X}.

\bibitem[Christov(2012)]{christov2012hidden}
Ivan~C Christov.
\newblock Hidden solitons in the {Zabusky--Kruskal} experiment: Analysis using
  the periodic, inverse scattering transform.
\newblock \emph{Mathematics and Computers in Simulation}, 82\penalty0
  (6):\penalty0 1069--1078, 2012.

\bibitem[Hirota(2004)]{hirota2004direct}
Ryogo Hirota.
\newblock \emph{The direct method in soliton theory}.
\newblock Number 155. Cambridge University Press, 2004.

\bibitem[Novikov et~al.(1984)Novikov, Manakov, Pitaevskii, and
  Zakharov]{novikov1984theory}
S~Novikov, Sergei~V Manakov, Lev~Petrovich Pitaevskii, and
  Vladimir~Evgenevi{\v{c}} Zakharov.
\newblock \emph{Theory of solitons: the inverse scattering method}.
\newblock Springer Science \& Business Media, 1984.

\bibitem[Osborne(2010)]{osborne2010nonlinear}
Alfred Osborne.
\newblock \emph{Nonlinear Ocean Waves and the Inverse Scattering Transform}.
\newblock Academic Press, 2010.

\bibitem[Lax(1968)]{lax1968integrals}
Peter~D Lax.
\newblock Integrals of nonlinear equations of evolution and solitary waves.
\newblock \emph{Communications on pure and applied mathematics}, 21\penalty0
  (5):\penalty0 467--490, 1968.

\bibitem[Magnus and Winkler(2013)]{magnus2013hill}
Wilhelm Magnus and Stanley Winkler.
\newblock \emph{Hill's equation}.
\newblock Courier Corporation, 2013.

\bibitem[Baddoo et~al.(2021)Baddoo, Herrmann, McKeon, Kutz, and
  Brunton]{baddoo2021physics}
Peter~J Baddoo, Benjamin Herrmann, Beverley~J McKeon, J~Nathan Kutz, and
  Steven~L Brunton.
\newblock Physics-informed dynamic mode decomposition (pidmd).
\newblock \emph{arXiv preprint arXiv:2112.04307}, 2021.

\end{thebibliography}

\end{document}